# A partial filament eruption in three steps induced by external magnetic reconnection

Jun Dai,[1, 2] Zhentong Li,[1, 2] Ya Wang,[1, 2] Zhe Xu,[1, 2] Yanjie Zhang,[1, 2] Leping Li,[3, 4] Qingmin Zhang,[1, 2] Yingna Su,[1, 2] and Haisheng Ji[1, 2]

[1]*Key Laboratory of Dark Matter and Space Astronomy, Purple Mountain Observatory, CAS, Nanjing, 210008,China*
[2]*School of Astronomy and Space Science, University of Science and Technology of China, Hefei, 230026,China*
[3]*Key Laboratory of Solar Activity, National Astronomical Observatories, CAS, Beijing 100101, China*
[4]*School of Astronomy and Space Science, University of Chinese Academy of Sciences, Beijing 100049, China*

## ABSTRACT

We present an investigation of partial filament eruption on 2012 June 17 in the active region NOAA 11504. For the first time, we observed the vertical splitting process during the partial eruption with high resolution narrow-band images at 10830 Å. The active filament was rooted in a small δ-sunspot of the active region. Particularly, it underwent the partial eruption in three steps, i.e. the precursor, the first eruption, and the second eruption, while the later two were associated with a C1.0 flare and a C3.9 flare, respectively. During the precursor, slow magnetic reconnection took place between the filament and the adjoining loops that also rooted in the δ-sunspot. The continuous reconnection not only caused the filament to split into three groups of threads vertically but also formed a new filament, which was growing and accompanied brightening took place around the site. Subsequently, the growing filament erupted together with one group splitted threads, resulted in the first eruption. At the beginning of the first eruption, a subsequent magnetic reconnection occurred between the erupting splitted threads and another ambient magnetic loop. After about three minutes, the second eruption occurred as a result of the eruption of two larger unstable filaments induced by the magnetic reconnection. The high-resolution observation provides a direct evidence that magnetic reconnection between filament and its ambient magnetic fields could induce the vertical splitting of the filament, resulting in partial eruption.

## 1. INTRODUCTION

Solar filaments are defined as the feature structures full of cool and dense plasma suspended in the solar corona (Engvold 1998; Labrosse et al. 2010; Mackay et al. 2010; Régnier et al. 2011; Parenti 2014; Karpen 2015; Gibson 2018), which are located along the magnetic po-

Corresponding author: Jun Dai
daijun@pmo.ac.cn

larity inversion lines (PILs) in the photosphere (van Ballegooijen & Martens 1989; Martin 1998; Parenti 2014; Gibson 2018). The high-resolution Hα observations have confirmed that filaments are composed of highly dynamic fine-scale threads (Chae et al. 2000; Lin et al. 2005, 2008; Schmieder et al. 2010; Berger 2014). The limb quiescent prominence threads appear to be long thick and predominately quasi-vertical (Berger & Haerendel 2009; Haerendel & Berger 2011), while the active filaments, located adjacent to sunspots (Engvold 2015), appear to be long thin and relatively horizontal (Okamoto et al. 2007), which may represent thin magnetic strings (Lin et al. 2005).

Generally, changes in the magnetic field topology of filaments could trigger eruptions, which includes tether-cutting (Moore et al. 2001) or flux cancellation (van Ballegooijen & Martens 1989), emerging magnetic flux (Chen & Shibata 2000), and magnetic flux injection (Chen 1996). Moreover, highly dynamic filaments could be eruptive due to the ideal magnetohydrodynamic (MHD) instabilities, including the torus instability (Kliem & Török 2006) and the ideal kink instability (Hood & Priest 1981; Török et al. 2004). Typically, filament eruptions are preceded by its precursor activities(Chen 2011), such as darkening and widening (Martin 1980), reconnection-favored emerging flux (Feynman & Martin 1995), large-amplitude oscillation (Chen et al. 2008; Zhang et al. 2012), heating and particle acceleration(Hernandez-Perez et al. 2019), and SXR brightening (Mitra et al. 2020). In addition, EUV or chromosphere brightenings inside the filament or its close vicinity are also considered as the precursors (Sterling & Moore 2005; Alexander et al. 2006; Sterling et al. 2011; Yan et al. 2013; Wang et al. 2018; Devi et al. 2020).

However, filament eruptions can be failed (Ji et al. 2003; Liu et al. 2009; Dai et al. 2021) result from the strong enough overlying magnetic fields (Török & Kliem 2005) or the lack of gained energy (Shen et al. 2011). The difference between full and failed eruptions is whether the filament mass and magnetic structure completely escaped from the Sun, producing the coronal mass ejections (CMEs, e.g. Dai et al. 2018), or not escaped at all. Occasionally, after experiencing a failed eruption and stopping at a certain height for hours, a filament would continue to erupt successfully, so called two-steps or multi-steps eruptions (Byrne et al. 2014; Gosain et al. 2016; Chandra et al. 2017; Filippov 2018). In addition, there is a kind of filament eruption that is called partial eruptions (Gilbert et al. 2007). In partial eruptions, filaments usually split into two parts with one part being fully erupted and another remained (Gilbert et al. 2000; Liu et al. 2007; Shen et al. 2012; Zheng et al. 2017; Cheng et al. 2018; Wang et al. 2020; Yang et al. 2020).

Typically, the graphic models (Gilbert et al. 2001) and the three-dimensional MHD simulations (Gibson & Fan 2006) believed that the magnetic reconnection within a filament leads to vertical splitting of the filament, and the extreme-ultraviolet (EUV) brightening can be detected at the splitting location (Tripathi et al. 2009; Shen et al. 2012; Tripathi et al. 2013; Zhang et al. 2015; Cheng et al. 2018). Meanwhile, Birn et al. (2006) suggested another mechanism which is called nonuniform magnetic twist through MHD simulations. Since a flux rope exists in places with different magnetic helicities, the twist of one part reaches kink instability and thus splits out and runs away, while the remaining part will have a low twist (Bi et al. 2015). Recently, Liu et al. (2012) found that partial eruptions could occur within the double-decker filaments, and only the upper filament erupted after being activated due to magnetic flux and current changes or the occurrence of magnetic reconnection(Liu et al. 2012; Kliem et al. 2014; Zheng et al. 2019; Pan et al. 2021).

Furthermore, by analyzing the 3D magnetic configuration of a partial eruption, Zhang et al. (2015) found that the runaway material may escape along the surrounding open magnetic fields, which suggested that magnetic structures around the filaments is also an influential factor to the partial eruption. Chen et al. (2018) reported that a filament splits into three parts on its ascent due to the tether-cutting (TC) reconnection below the middle of the filament, and only the middle high-lying part erupted. More recently, Monga et al. (2021) highlighted that the reconnection process between filament magnetic field and its ambient loops can induce the splitting of the filament, which is supported by the similar results from high-resolution observation (Li et al. 2016; Xue et al. 2016). These studies have provided new constraints to partial filament eruptions, but the specific mechanisms still need more detailed analysis.

On 2012 June 17, a partial filament eruption occurred in the active region 11504 were well recorded with He I 10830 Å narrow-band images by the Goode Solar Telescope (GST) at Big Bear Solar Observatory (BBSO).

The high-resolution observation gives us a good opportunity to analyze and understand the mechanisms of partial eruptions. Particularly, the filament underwent the partial eruption in three steps, which was associated with two C-class flares and multiple hot channels. Zeng et al. (2014) have reported excitation of He I atoms by EUV illumination by analyzing the emission flux of the flares from the corona to the transition region and then to the chromosphere. In this paper, we focus on the triggering process and dynamic behaviour of the partial eruption. The observation and data analysis are described in Section 2. The results are presented in Section 3. The detailed interpretation and discussion are presented in Section 4 and a brief summary are given in Section 5

## 2. OBSERVATION AND DATA ANALYSIS

The New Solar Telescope (NST) is a 1.6 m aperture ground-based off-axis telescope at Big Bear Solar Observatory (BBSO; Goode et al. 2010), which was renamed as the Goode Solar Telescope (GST) in 2017 July. Since the off-axis design was removed central obscuration, GST can significantly reduce stray light, and long duration good seeing conditions combined with the high-order Adaptive Optics (AO) system enable to provide consecutive observations for solor activities with diffraction-limited images (Cao et al. 2010a). On 2012 June 17, from 17:01 UT to 18:08 UT, GST pointed to the active region NOAA 11504, targeting a small active filament. High-resolution filtergrams were obtained in narrow-band (NB) Lyot filters in He I 10830 Å blue wing (0.25 Å, bandpass: 0.5 Å) and in H$\alpha$ 6563 Å blue wing (-0.75 Å, bandpass: 0.25 Å), and also in a broadband filter (bandpass: 10 Å) containing the TiO 7057 Å line. The 10830 Å data were acquired by employing the high sensitivity HgCdTe CMOS IR focal plane array camera (Cao et al. 2010b), with a cadence of 10s, pixel size of $0''.0875$ with a field of view (FOV) of $90'' \times 90''$. The H$\alpha$ image scale is $0''.03$ pixel$^{-1}$ and has a cadence of 30s, with a FOV of $60'' \times 60''$. The TiO images have a spatial sampling and time cadence of $0''.0375$ pixel$^{-1}$ and 30s, respectively, with a FOV of $70'' \times 70''$.

The full-disk photospheric continuum intensity images and line-of-sight (LOS) magnetograms were observed by the Helioseismic and Magnetic Imager (HMI; Schou et al. 2012) on board Solar Dynamics Observatory (SDO, Pesnell et al. 2012) with a spatial resolution of $1''.2$ and time cadence of 45 s. HMI also provides the continuous vector magnetic fields (Turmon et al. 2010) in the HMI Active Region Patches (HARPs) region. In addition, we employ UV and EUV images from the Atmospheric Imaging Assembly (AIA, Lemen et al. 2012) on board SDO. For level 1.5 AIA images, their spatial sampling is fixed at $0.6''$ pixel$^{-1}$ and have a time cadence of 12s. The level 1 data from AIA and HMI were calibrat- ed using the standard Solar SoftWare (SSW) programs aia_prep.pro and hmi_prep.pro, respectively. Soft X- ray (SXR) light curves of the flare were recorded by the GOES spacecraft with a cadence of 2s. The associat- ed CMEs were observed by the C2 on board the SOHO Large Angle and Spectrometric Coronagraph (LASCO; Brueckner et al. 1995).

The two large sunspots and the small $\delta$ sunspot are recognizable in HMI intensity images, which are very helpful for a convincing co-alignment between the local TiO and full-disk intensity images. The HMI intensity and TiO images were co-aligned by carrying out an automatic mapping approach developed by Ji et al. (2019). Visual inspections and error corrections were made during the co-alignment,and we estimate the accuracy to be within $0.5''$.

In order to reconstruct the coronal magnetic fields before eruptions in the AR 11504, we apply the nonlinear force-free field (NLFFF; Wiegelmann et al. 2006, 2012) model using the magneto-frictional method (Guo et al. 2016), which is implemented in Message Passing Interface Adaptive Mesh Refinement Versatile Advection Code (MPI-AMRVAC; Keppens et al. 2003, 2012; Porth et al. 2014). The calculation is performed within a box of $288 \times 170 \times 170$ uniform grid points, which almost covers the whole AR.

## 3. RESULT

### 3.1. Overview

As shown in Figure 1(a-b), there were three main sunspots (one leading and two following) located in the active region 11504. The negative sunspot in the southeast and the positive one in the west are labeled as N1

4 Dai et al.

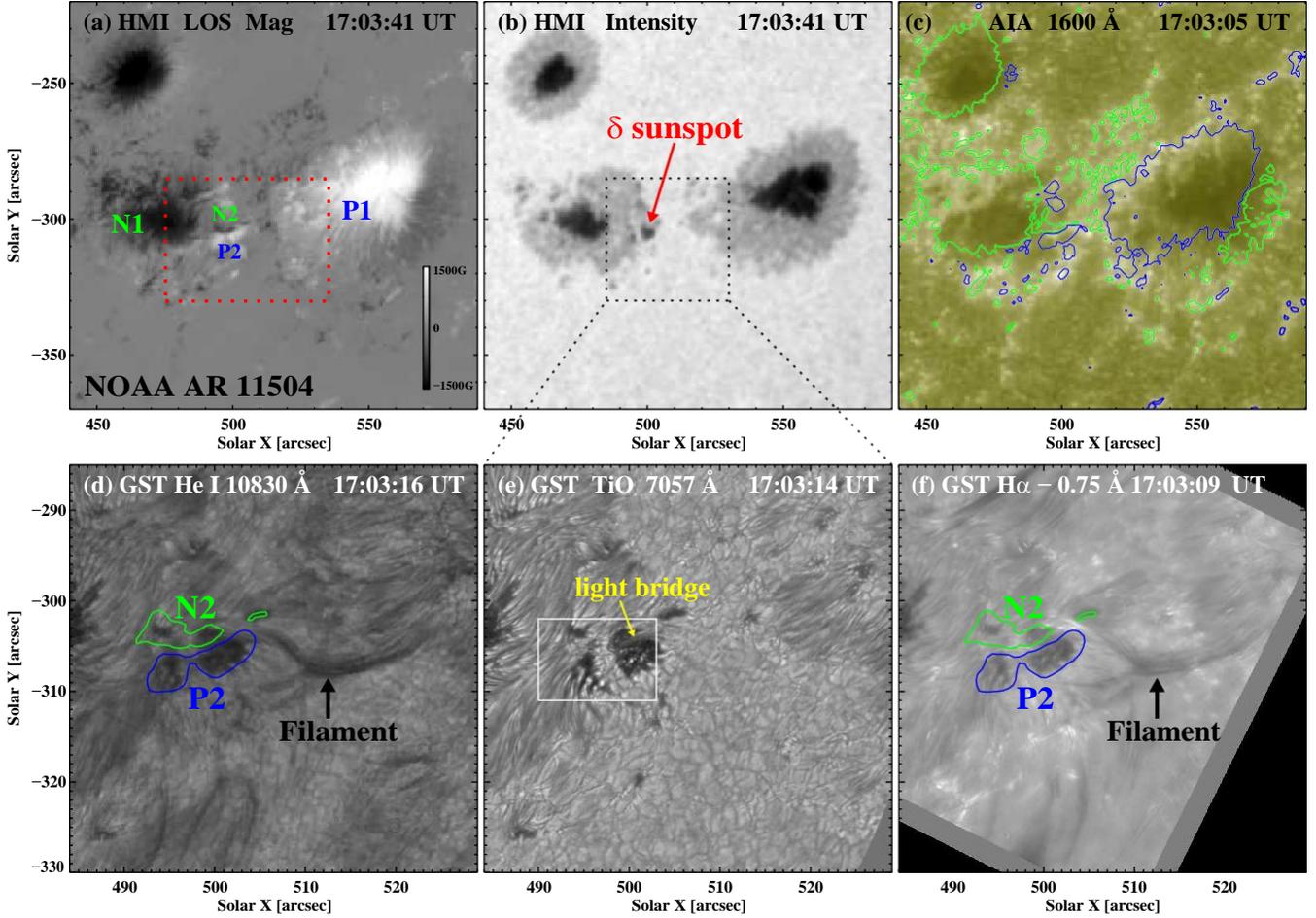

**Figure 1.** Overview of the active region 11504 including the filament and the $\delta$ sunspot. Panels (a-c) give a LOS magnetogram and an intensity image observed by HMI and a UV image in 1600 Å observed by AIA. Panels (d-f) show enlarged images from the boxed area in panel (b) of He I 10830 Å, TiO 7057 Å, and H$\alpha$ blue wing observed by GST. The red dotted rectangle in (a) gives the FOV of Figure 3 and Figure 4(a-f). The negative and positive magnetic fields of sunspots are denoted by N1,N2,P1,P2 in panel (a). The green and blue contours in (c) represent the negative and positive magnetic fields with the magnetic strengths of -200 and 200 G. The negative and positive part of the $\delta$ sunspot are denoted by green and blue contours with the magnetic strengths of -250 and 450 G in panels (d) and (f). The total flux of 10830 Å and TiO intensity, also the total positive and negative magnetic fluxes within the white box in (e) are calculated and their temporal profiles are plotted in Figure 2(b-c).

and P1, respectively. At ∼17:00 UT, a small $\delta$ sunspot was formed between N1 and P1 with two magnetic polarities as labeled with N2 and P2. The bottom panels in Figure 1 show further local high-resolution images, from which we can see that there was a light bridge inside the $\delta$ sunspot, the position corresponds to the magnetic interface between N2 and P2. In addition, for the filament we will analyze, the left end is rooted between the polarities N2 and P2. However, the magnetic polarity inversion line associated with the filament can be clearly recognized only in the $\delta$ sunspot.

Figure 2(a) gives the SXR light curves, showing the C1.0 flare and the C3.9 flare in 1-8 Å (red line) and 0.5-4 Å (magenta line). To investigate the evolution of radiation and magnetic flux for the $\delta$ sunspot during the filament eruption, we calculated the total flux (normalized) of 10830 Å and TiO intensity, also the total positive and negative magnetic fluxes within the white box in Figure 1(e). Their temporal evolution is plotted in Figure 2(b-c). In particular, as can be seen from the red curve in Figure 2(b) for 10830 Å, the filament eruption underwent three processes: precursor, the first eruption and the second eruption. The two eruptions corresponded to the two flares in Figure 2(a), first with the C1.0 flare at ∼17:27 UT, and then with the C3.9 flare at ∼17:39 UT. It is noteworthy that the radiation in 10830 Å had a slow increase including a small peak at ∼17:20 UT. Visual inspections to the movies of Hα images show continual brightening around the filament. From 17:05 UT to 17:25 UT, the TiO intensity continued to increase slowly, and the positive flux of the δ sunspot continued to decrease, while the negative flux begin to decrease after the C1.0 flare, which indicates the occurrence of the magnetic cancellation or reconnection before and during the first eruption.

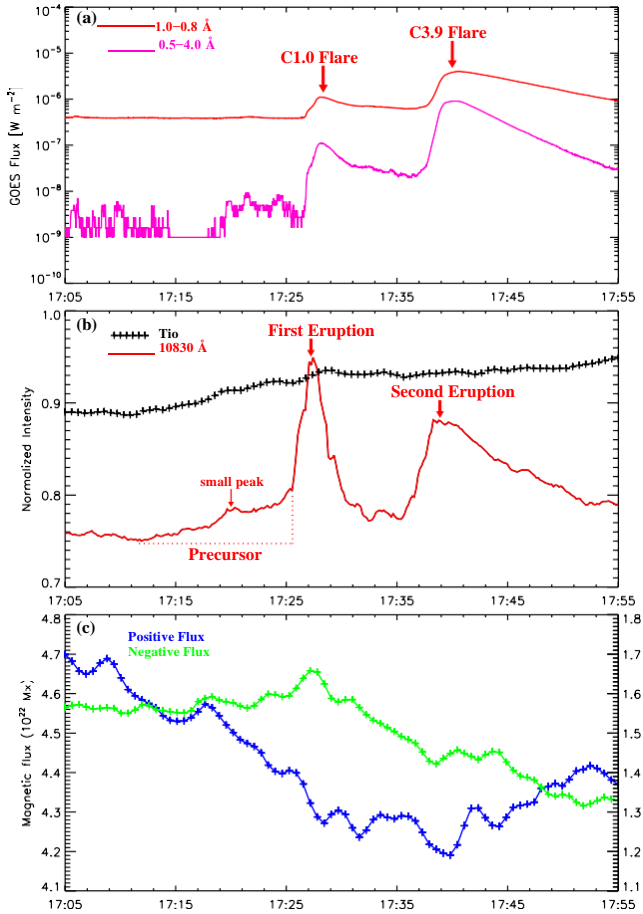

**Figure 2.** (a) SXR light curves of the two flares in 1-8 Å (red line) and 0.5-4 Å (magenta line). (b) Light curves (normalized) in 10830 Å and TiO within the white box of Figure 1(e). (c) Temporal evolutions of the total positive and negative magnetic fluxes within the white box of Figure 1(e). An animation showing the light curves and magnetic flux in the delta sunspot region in panels (b) and (c) is available. These are show on the right side of the animation. The left side of the animation shows the He I 10830 Å (top; see also panels (a1-a4) of Figure 3), TiO (middle), and LOS-mag (bottom) images. The animation realtime duration is 12 seconds which covers 50 minutes starting at 17:05 UT and ending the same day at 17:55 UT.

### 3.2. The precursor phase

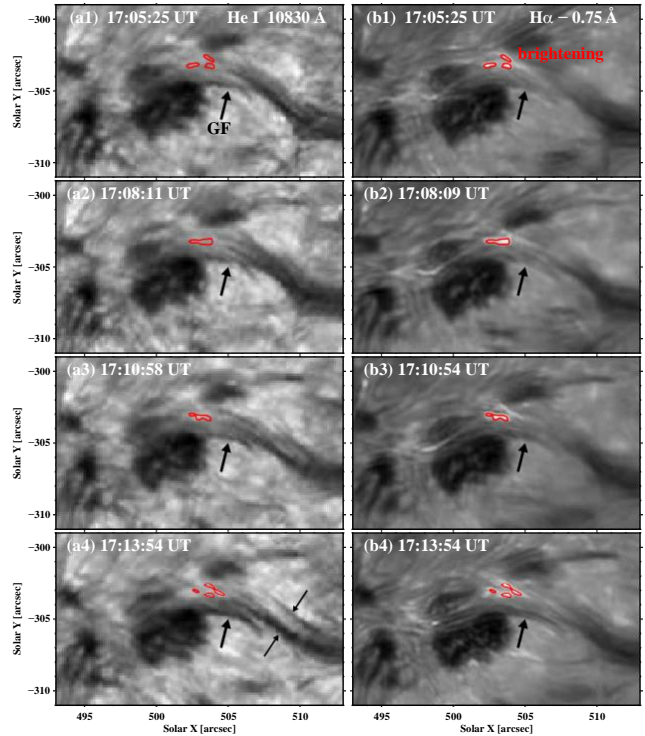

**Figure 3.** The growing process of the filament produced by magnetic reconnection. Left panels show four images in He I 10830 Å (a1-a4) and right panels show four images in Hα blue wing (b1-b4). GF is pointed by black thick arrows. The brightening near N2 is enclosed by red contours. The spitting threads was marked with black thin arrows in panel (a4). An animation of the Hα blue wing (panels b1-b4) observations is available, it begins at 17:03 UT and ends the same day at 17:16 UT. The animation realtime duration is 4 seconds and it shows continuous brightening near to the eastern leg of the Filament. The He I 10830 Å sequences is available in the top right portion of the animation available with Figure 2.

Figure 3 presents local high-resolution images for the δ sunspot before the precursor in 10830 Å (left panels) and Hα blue wing (right panels). At ∼17:03 UT, when we started to observe, there was already a growing filament (GF) near the footpoint of the Filament, which is pointed by thick black arrows. The GF was gradually growing with continuous brightening closed to the eastern leg of the Filament connected to N2, which is enclosed by red contours. Starting around 17:13 UT, the eastern leg of the filament began to bifurcate into two groups of fine threads, which is marked with two thin black arrows in Figure 3(a4). Until ~17:24 UT, as shown in Figure 4(g), there were three groups of discernible threads, which are designated as Threads 1 (TH1), Threads 2 (TH2), Threads 3 (TH3).

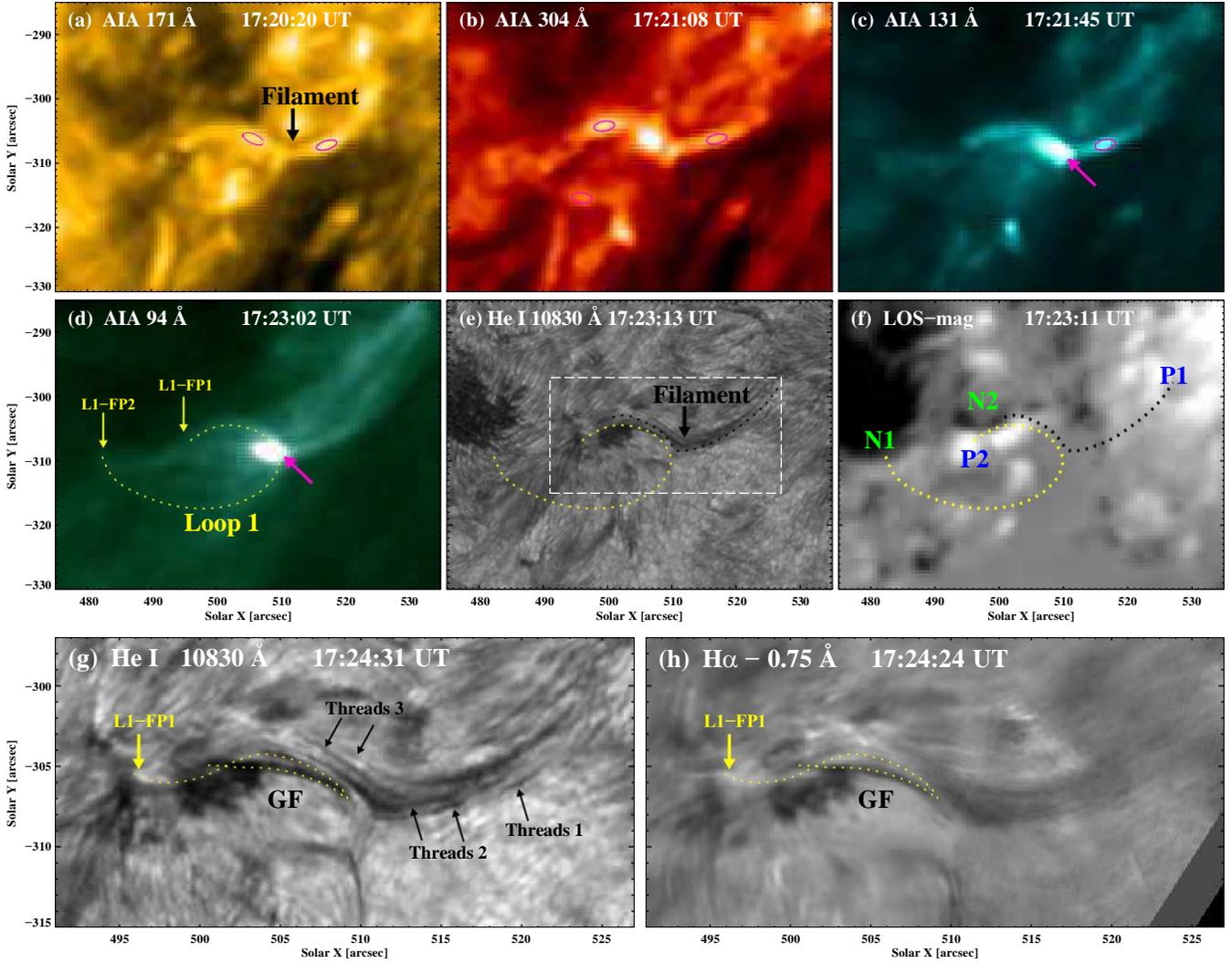

**Figure 4.** The cusp-type formed between Threads 1 in the filament and Loop 1. Panels (a-d) show EUV images in 171 Å, 304 Å, 131 Å, 94 Å. Other panels give two 10830 Å narrow band filtergrams (e,g), an HMI LOS magnetogram (f) and an Hα blue wing image (h). The white rectangle in (e) gives the field of view (FOV) of (g-h). The outflow and reconnection point are enclosed (pointed) by the magenta ellipses (arrows) in panels (a-d). Loop 1 is outlined by yellow dashed lines in panel (d-f) and the two foot-points of Loop 1 (L1-FP1,L1-FP2) is pointed by yellow arrows in panels d,g and h. Threads 1 is outlined by the black dashed lines in panels (e-f). The negative and positive magnetic fields are separately denoted by N1,N2,P1,P2 in the panel (f). The growing filament (GF) is outlined by the yellow dashed lines in panels (g-h). An animation of the top 6 panels is available, it begins 17:05 UT and ends the same day at 17:25 UT, lasting 5 seconds.

Figure 4 shows a cusp-type configuration formed between the filament and its adjoining loops (labeled as Loop 1 in Figure 4(d)). Starting around 17:20 UT, the bright features, presumably plasmoid structures flowing along the filament and Loop 1 were detected in EUV bands, which were enclosed by the magenta ellipses in Figure 4(a-c). Almost simultaneously, a persistent reconnection point, which is marked by the magenta arrows in Figure 4(c-d), was observed at the interface between the filament and Loop 1 in 131 Å and 94 Å. The filament and Loop 1 were separately denoted by black and yellow dotted lines in Figure 4(d-e). The corresponding magnetic fields are given in Figure 4(f). It is obvious that two foot-points of Loop 1 (labeled as L1-FP1 and L1-FP2) were rooted in P2 and N1 while the filament was rooted in P1 and N2. In Figure 4(g-h), it can be seen that one foot-point of the GF was located at P2 (i.e. previous L1-FP1), while another foot-point rooted in N2 (i.e. negative foot-point of the filament). In addition, the NLFFF extrapolation is performed to investigative the 3D magnetic fields in the source region, using the HMI vector magnetograms before eruptions (17:12 UT), and the extrapolated results are present in the top and side view in Figure 5. Obviously, there are a flux rope (red lines) and a group of magnetic loop (yellow lines), which nearly correspond to the filament and Loop1 in Figure 4(d-e), and their foot-points are consistent with those in the Figure 4(f).

### 3.3. The first eruption

Figure 6 presents the process of the first eruption and the interaction between Th2 and its nearby magnetic structures. Starting around 17:25 UT, the GF became unstable and erupted together with TH2, associated with a C1.0 class flare.

The erupting TH2 and the remaining TH3, as well as their footpoints, were marked with red arrows and letters in Figure 6(a2). The separated footpoints indicate that TH2 and TH3 have been completely splitted.

Subsequently, at ∼17:26 UT, the erupted GF was observed in the 94 Å (see Figure 6(c1-c2)), while another group of loops (labeled as Loop 2) appeared, with two footpoints can be designated as L2-FP1 and L2-FP2. As shown in Figure 6(b-c), at the interface of Th2 and Loop 2, there was a continuous brightening in 94 Å and 131 Å and it was accompanied by the outflow along Loop 2. However, the outflow only appeared an absorption feature in 10830 Å. These significant signals indicate that TH2 and Loop 2 underwent another magnetic reconnection. Of course, there is another possibility that the merging or reconnection between different threads of the filaments, including TH2, GF and F3 (present in

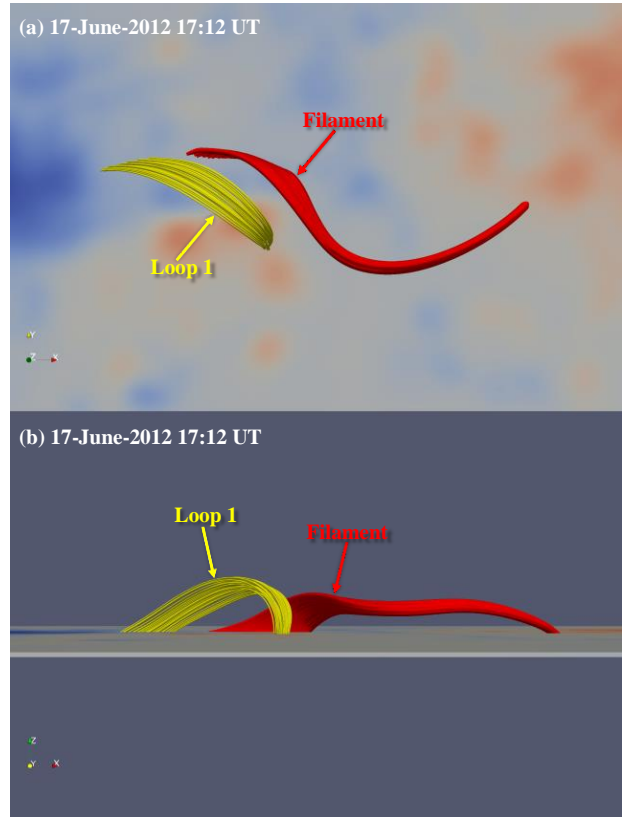

**Figure 5.** Top(top panel) and side(bottom panel) views of the NLFFF extrapolated Filament(red) and Loop 1(yellow) at 17:12 UT on 2012 June 17. The FOV of these panels is approximated to the FOV of Figure 4(f).

Figure 7(a)). It is worth noting that the GF was above the brightening feature, indicating that it is higher.

One distal end of the initial flare ribbon was pointed by yellow arrows (L1-FP1) in Figure 6. Around 17:27 UT, the flare ribbon elongated to L2-FP1, and the radiation was significantly enhanced, corresponding to the peak in 10830 Å for the first eruption (see Figure 2(b)). Concurrently, there appeared an another brightening region of the flare ribbon, which is encircled in yellow circles in Figure 6(b3,c3,d). The brightening region is found to be one foot-point of a hot channel formed after the eruption, while another footpoint is encircled with red circles. As shown in Figure 6(d1), both footpoints of the hot channel are rooted in the flare ribbon.

### 3.4. The second eruption

Figure 7 presents the process of the second eruption. From 17:31 UT to 17:34 UT, as shown in Figure 7(a),

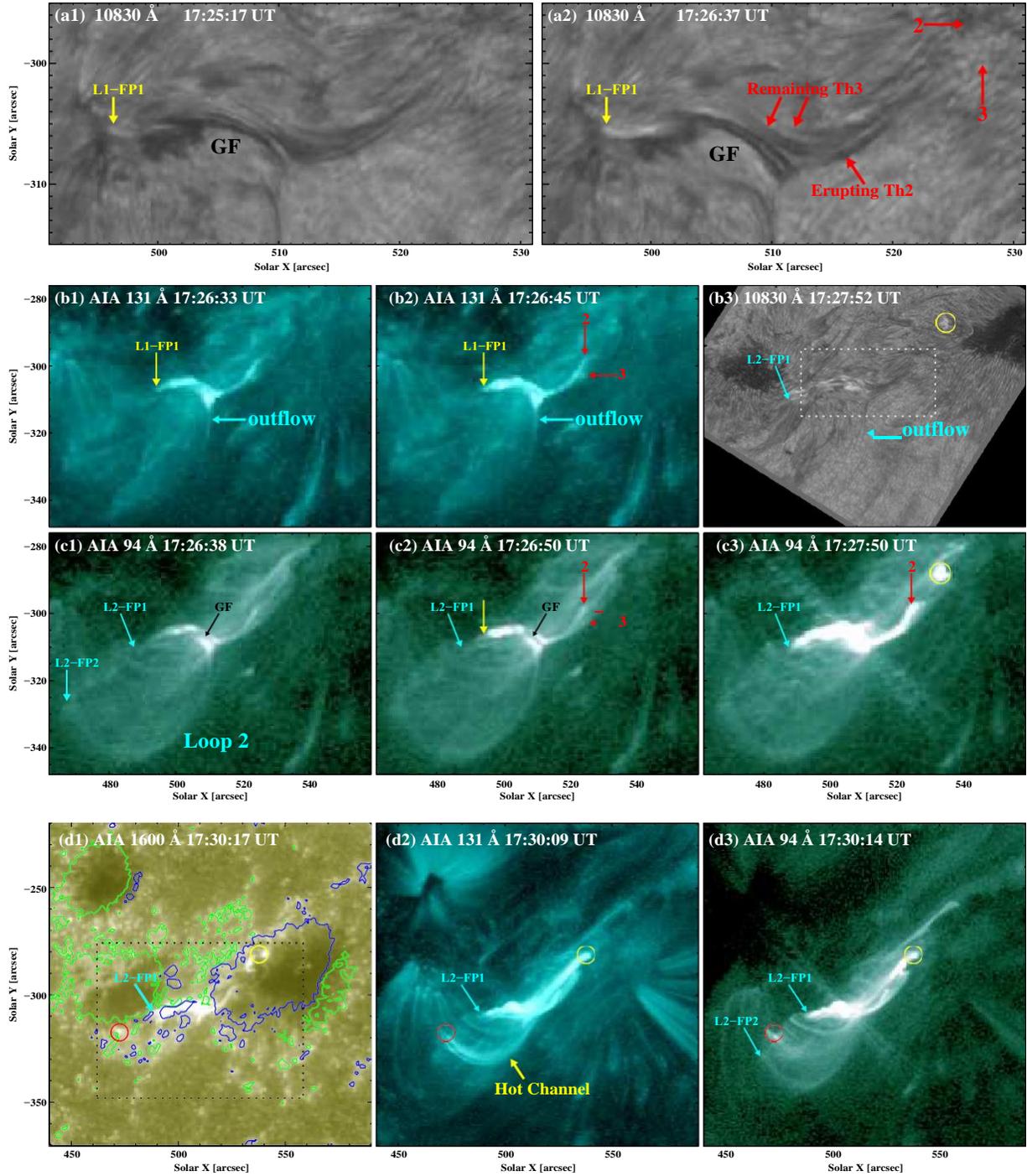

**Figure 6.** The process of the first eruption and the interaction between Th2 and its nearby magnetic structures. Panels a1,a2 and b3 give three 10830 Å narrow band filtergrams and other panels show images in AIA 131 (Å) (b1,b2 and d2), 94 Å (c1,c3 and d3) and 1600 Å (d1). The white dashed rectangle in (b3) gives the field of view (FOV) of (a1-a2) and the black dashed rectangle in (d1) gives the field of view (FOV) of (b-c). GF is marked with black letters and arrows in panels (a-c). The Erupting threads 2 and Remaining threads 3 are pointed by red arrows in (a2). The footpoints of the Erupting threads 2 and Remaining threads 3 are separately marked with red letters 2 and 3 in panel (a-c). The outflow between Th2 and Loop 2 is marked with cyan thick arrows in panels (b1-b3). The two footpoints of the Loop 2 (L2-FP1, L2-FP2) are marked with cyan thin arrows in (b3,c1-c3, and d1-d3). The yellow and red circles in (b3,c3 and d1-d3) enclosed the initial flares. The green and blue contours in (d1) is same to that in Figure 1. An animation of the panels (b-c) is available, it begins at 17:25 UT and ends the same day at 17:30 UT, lasting 6 seconds, which shows the first eruption.

there was no significant brightening in 10830 Å and H$\alpha$ blue wing, corresponding to the valley between the first and second eruption on the 10830 Å time profile (see Figure 2(b)).

At about 17:31 UT, another filament (labeled as F3) formed by the reconnection between Loop 1 and Th1 is detected in 10830 Å and H$\alpha$ blue wing, which is outlined by the yellow dashed line in Figure 7(a1,a3). The two footpoints of F3 were pointed by yellow and red arrows, and corresponded to the previous L1-FP2 and positive footpoints of TH1, respectively. Similarly, a filament (labeled as F4) was also detected in 10830 Å and H$\alpha$ blue wing, which was marked with cyan and red arrows together in Figure 7(a). As described in the previous section, F4 may be formed by the reconnection between Loop 2 and Th2, and the possibility of the merging of different filaments cannot be excluded. Both F3 and F4 were unstable and erupted at ∼17:37 UT, associated with a C3.9 class flare and another weak CME as shown in Figure 8(a3). In Figure 7(b), it shows multiple initial brightenings shown as the flare ribbons and the remaining filament after two eruptions. Two distal ends of the initial flare ribbon were denoted by yellow arrows and circles. Particularly, the positive footpoint of F3 (pointed by a red arrow) did not appear brightening in 10830 Å, but there was a significant brightening in the EUV bands. Similar to the first eruption, the footpoints of the hot channel formed during the second eruption were also rooted in the flare ribbon but in different regions, where were the positive footpoint of F3 and the elongation region of the initial flare ribbon (enclosed by red circles). This suggested that multiple hot channels were formed by the two eruptions. At about 17:40 UT, the flare radiation reaches its highest, corresponding to the curve peak for the second eruption in 10830 Å (see Figure 2(b)). The post-flare loops were outlined by the black contours in Figure 7(c2,c3).

## 4. INTERPRETATION AND DISCUSSION

Based on the observation, schematic diagrams showing the evolution of the partial eruption are provided in Figure 9. As shown in the enlarged partial diagram in Figure 9(a), Loop 1 was composed of thin magnetic field lines, while the filament was composed of fine threads. The group of threads that reconnected with Loop 1 was marked as TH1, which was observed in Figure 4(g). When Loop 1 and the filament approached each other, the magnetic reconnection started. The magnetic lines in Loop 1 and the threads of the filament were disconnected at the position marked by the asterisk. Then the eastern part of TH1 reconnected to the western leg of Loop 1, forming the GF, while the western part of TH1 reconnected to the eastern leg of Loop 1, forming another new filament F3. The continuous accumulation of new magnetic lines corresponds to the progressive growth of the GF. Apart from TH1, the remaining threads in the filament split into two groups (i.e. TH2 and TH3) vertically during the reconnection between TH1 and Loop 1. Then, the GF erupted together with Th2. Magnetic reconnection taking place between the erupting Th2 and Loop 2, or the merging interaction between different threads of the filaments, produced another unstable filament F4, although only the former case is shown here. Finally, F3 and F4 erupted together and Th3 was remained. As shown in Figure 8, the associated weak outflow was detected by LASCO/C2.

The three-step eruptions we report here have hardly been reported before. Earlier, multi-step eruptions were when a filament experienced a failed eruption and then erupted successfully, and the interval between eruptions was at least one hour or longer (Byrne et al. 2014; Gosain et al. 2016; Chandra et al. 2017). Filippov (2018) proposed the existence of two critical heights as a result of the non-monotonic coronal magnetic field. Therefore we presume that such critical heights should also exist for the active region here, this may explain why F3 did not erupt with GF together, but only with F4 after the subsequent interaction.

In this study, the filament and its nearby loops (Loop 1 and Loop 2) was rooted in a same $\delta$ sunspot. Figure 10 present the vector magnetic fields of the $\delta$ sunspot region, which originate from the negative and positive vertical fields. Obviously, the magnetic fields in the $\delta$ sunspot region were more compact with opposite polarities, which were further identified as high-gradient, strongly-sheared and non-potential (Hagyard et al. 1984; Tanaka 1991; Schrijver 2007). Such non-potential fields tend to produce the magnetic reconnections and eruptions in case of photospheric motions (Fang & Fan 2015; Liu et al. 2021). Hence, the filament and its nearby loops were favored configurations for the magnetic reconnections and eruptions. In this case, the reconnection of the filament (Threads 2) and its nearby loops (Loop 2) may still be occurring during the first eruption. However, the observation is not very clear from the present images. Therefore, the continuous brightening during the first eruption in 131 Å and 94 Å could also result from the merging of TH2, GF and F3 in the favored magnetic configurations (Schmieder et al. 2004; DeVore et al. 2005; Török et al. 2011; Chandra et al. 2011).

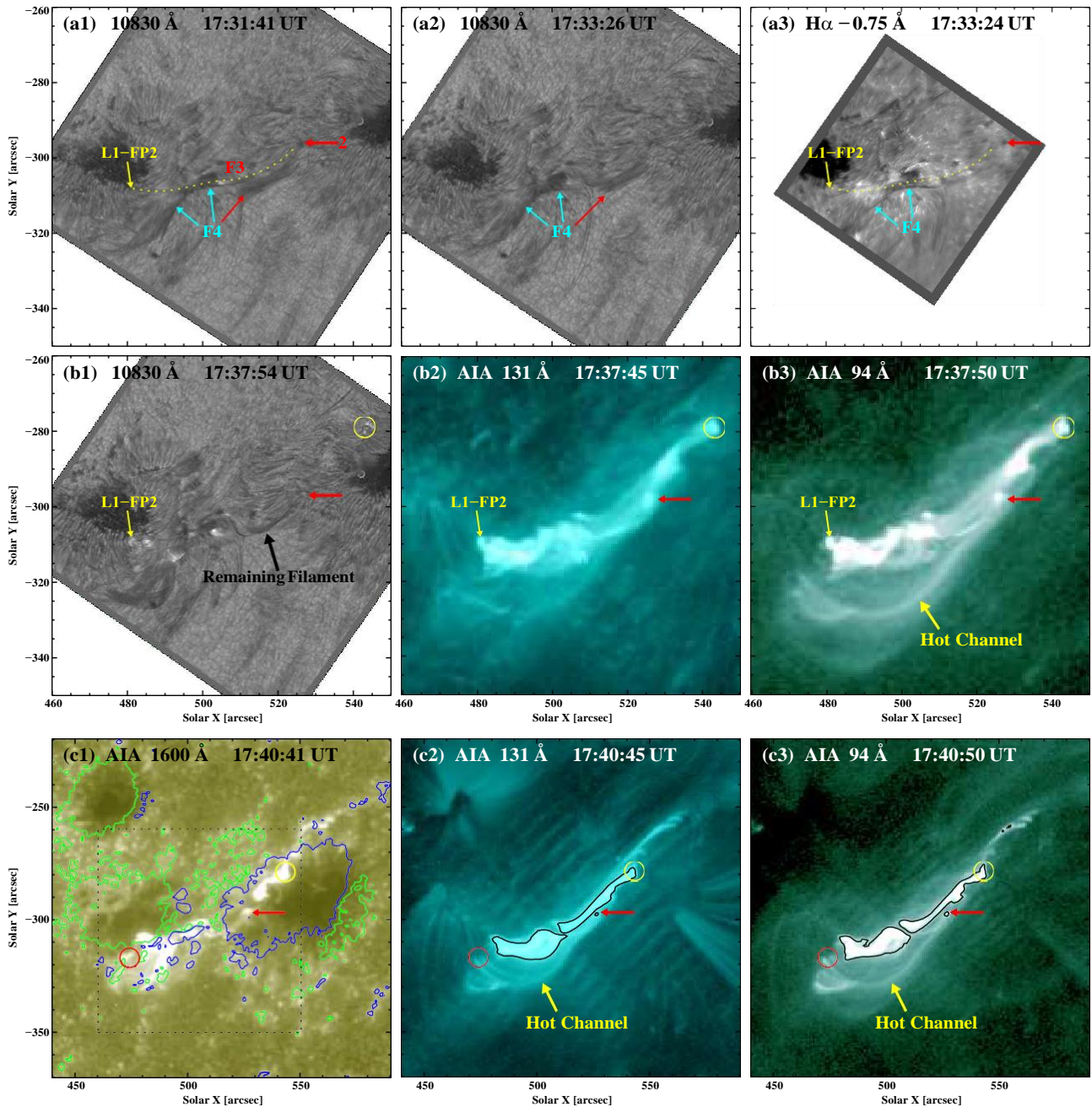

**Figure 7.** The process of the second eruption. Panels a1,a2 and b1 give three 10830 Å narrow band filtergrams and panel a3 shows an Hα blue wing image. Other panels show images in AIA 131 Å (b2 and c2), 94 Å (b3 and c3), and 1600 Å (c1). The black dashed rectangle in (c1) gives the field of view (FOV) of (a-b). Filament 3(F3) is outlined by the yellow dashed lines while the cyan and red thin arrows in (a) point to Filament 4(F4). The footpoints of F3 (e.i. the footpoint of Erupting Th2 and L1-FP2) are separately marked with red thick arrows and yellow thin arrows in (a-c). The initial flares are enclosed by the yellow and red circles in (b-c) . The green and blue contours in (c1) is same to that in Figure 1 . The black contours in (c2-c3) outline the post-flare loops. An animation of the panels (b1-b3) is available, it begins at 17:30 UT and ends the same day at 17:55 UT, lasting 6 seconds, which shows the second eruption.

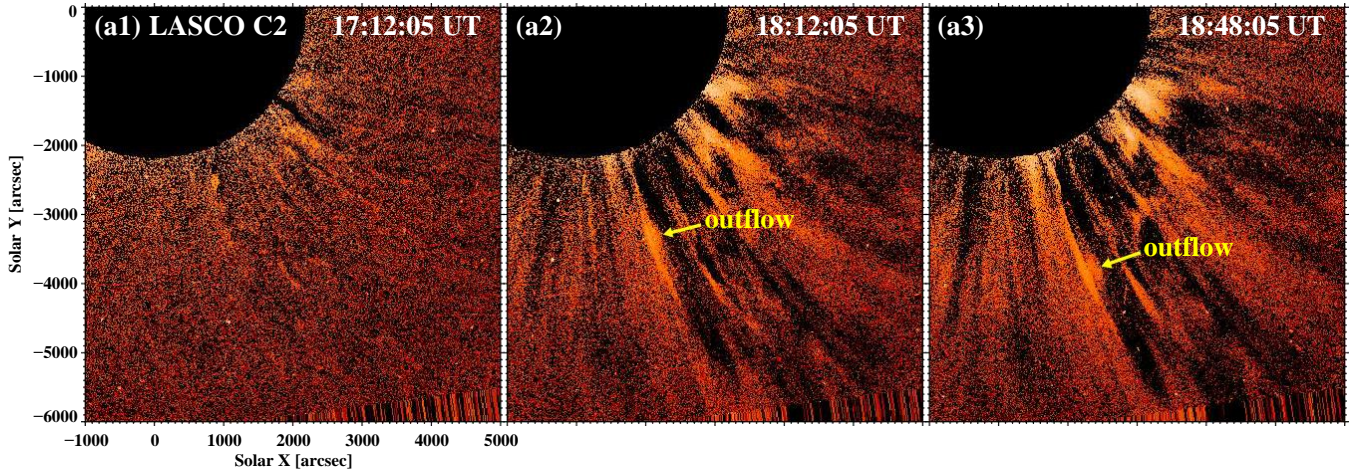

**Figure 8.** Three snapshots of the white-light CME observed by LASCO/C2. An animation of this figure is available, it begins at 17:12 UT and ends the same day at 19:48 UT, lasting 3 seconds.

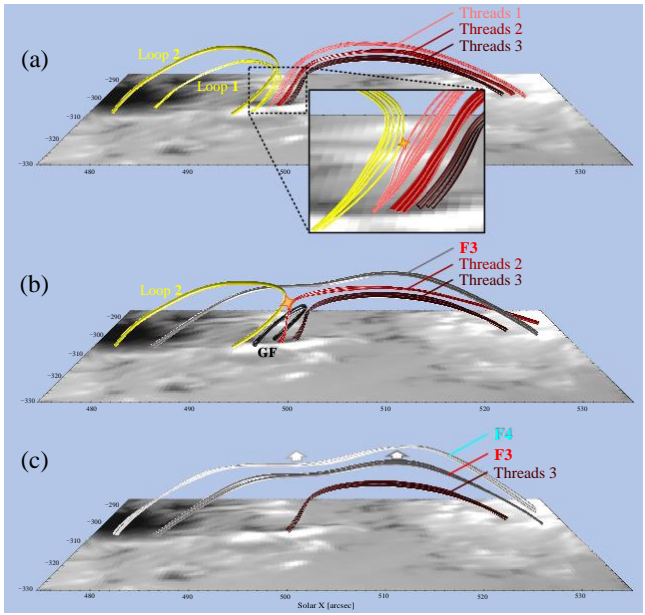

**Figure 9.** Schematic diagrams showing the scenario of vertical splitting in the partial eruption. The yellow lines in panels (a) and (b) represent Loop 1 and Loop 2. The red lines with different color depths represent the threads of the filament. The black, dark grey and light grey lines separately represent GF, F3 and F4. The reconnection sites are denoted by star symbols.

Usually, the precursors of brightening prior to filament eruptions are closely related to the triggering mechanism for eruptions. Typically, transient X-ray brightenings, so-called preflares, are observed before the flares (Chifor et al. 2007; Sterling et al. 2011), which are coincident with magnetic cancellation or emerging magnetic flux (Schmieder et al. 2008). Moreover, Yan et al. (2013) suggested that the brightening in the chromosphere before the filament expansion may indicate the onset of TC reconnection. And Chen et al. (2018) detected obvious brightenings in UV bands and hot outflows in the chromosphere, which showed the preflare reconnection process of the partial eruptions. Similarly, we also found continuous chromosphere brightening near the foot-point of the filament and magnetic cancellation during the growth of the GF. Therefore, we believe that the magnetic reconnection between the filament and Loop 1 has already begun when the GF appears. The magnetic reconnection continued until the brightening plasmoids and CS appeared in EUV bands. We support that the reconnection process was a transition from the low corona to the high corona, hence no characteristic signals in EUV bands was detected at the beginning.

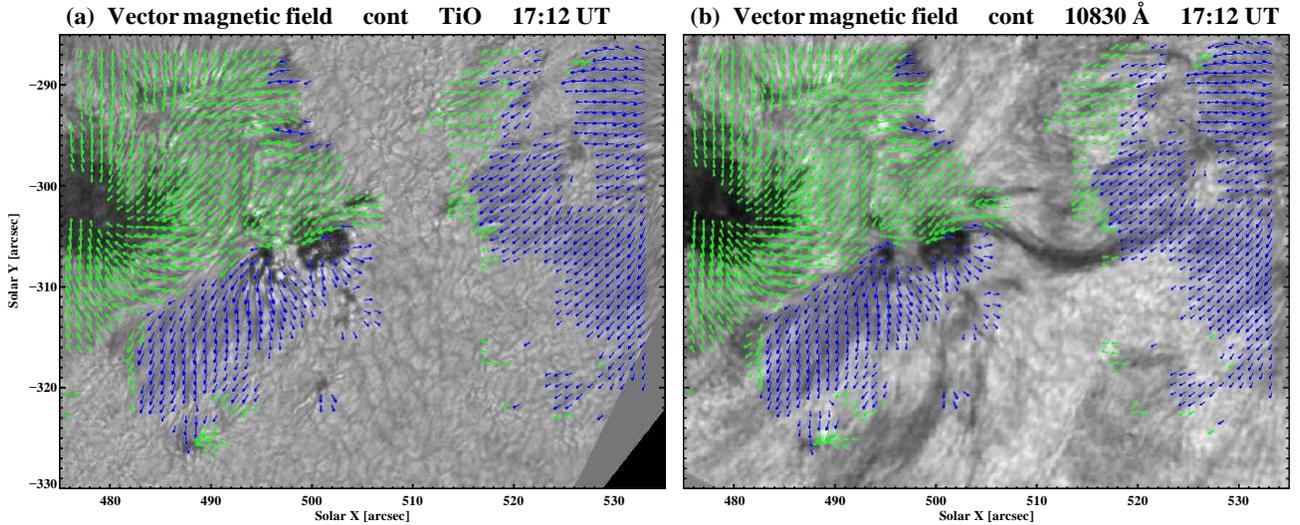

**Figure 10.** Horizontal magnetic fields, which originate from the areas of the negative and positive vertical magnetic fields, are separately depicted by green and blue arrows, and overlaid on the simultaneous TiO and 10830 Å images at 17:12 UT.

Generally, the splitting is mostly caused by internal magnetic reconnection (Gibson & Fan 2006), which tends to take place in case of shearing motions or so called filament-hosting flux rope (Cheng et al. 2018). However, Chen et al. (2018) present a three-part splitting due to tether-cutting reconnection below the filament and Zheng et al. (2019) highlighted the role of the TC reconnection between the overlying sheared loops. Meanwhile, Monga et al. (2021) proposed that the reconnection between a filament and its nearby EUV loops could induce the vertical splitting through destabilizing the magnetic topology of the filament, hence we believe that the splitting found in our observation was also caused by a similar external magnetic reconnection. Such interaction during the reconnection may be the transfer of magnetic flux between the filament and its nearby loops (Kliem et al. 2014). Thus, we support that the eruption of TH2 was due to the magnetic flux injection from Loop 1. In conclusion, magnetic reconnection between a filament and its nearby loops could induce the splitting and partial eruption of the filament.

In addition, although the footpoints of the hot channels produced during two eruptions were root in the flare ribbon, but they were not identical, which implied the existence of multiple hot channels and suggested that the forming of hot channels was closely associated with flares. The hot channels in this event are consistent with the characteristics summarized by Cheng & Ding (2016), i.e. EUV brightening structures with the high temperature rooted in the penumbra or penumbra edge. However, the hot channels here are not sigmoidal but arcaded.

## 5. SUMMARY

In this paper, we reported a three-step partial eruption of an active filament observed by GST in AR 11504 on 2012 July 17. With the help of high-resolution imaging at 10830 Å, we directly observed the vertical splitting process. Combined with the multi-wavelength data from SDO/AIA and HMI, we investigated detailed process of the precursor and two successive eruptions. We confirm that the vertical splitting and partial eruptions were induced by the magnetic reconnection between the filament and its nearby loops, which is so-called external magnetic reconnection. Its initial stage went on gently, giving rise to the slowly growing filament and the slowly rising brightening, which constitute the precursor phase of the partial eruptions. Such a precursor was totally missed in the GOES light curves. In conclusion, such a kind of external magnetic reconnection should be a potential triggering mechanism for partial filament eruptions.

## 6. ACKNOWLEDGEMENTS

The authors thank Dr. Yijun Hou in National Astronomical Observatory for constructive suggestions. SDO is a mission of NASA's Living With a Star Program. AIA and HMI data are courtesy of the NASA/SDO science teams. BBSO operation is supported by NJIT and US NSF AGS 1821294 grant. GST operation is partly supported by the Korea Astronomy and Space Science Institute, the Seoul National University, and the Key Laboratory of Solar Activities of Chinese Academy of Sciences (CAS) and the Operation, Maintenance and Upgrading Fund of CAS for Astronomical Telescopes and Facility Instruments. Y.W. is supported by the Youth Fund of JiangSu (No. BK20191108). Q.M.Z. is supported by the CAS Key Laboratory of Solar Activity, National Astronomical Observatories (KLSA202006) and the International Cooperation and Interchange Program (11961131002). This work is supported by the National Natural Science Foundation of China grant NO. 41761134088, 11790302 (11790300), 12073042, 12003072, 11773079, 11473071, and U1731241 as well as the Strategic Priority Research Program on Space Science, CAS, Grant No. XDA15052200 and XDA15320301.

## REFERENCES

Alexander, D., Liu, R., & Gilbert, H. R. 2006, ApJ, 653, 719. doi:10.1086/508137

Berger, T. E. & Haerendel, G. 2009, AGU Fall Meeting Abstracts

Berger, T. 2014, Nature of Prominences and their Role in Space Weather, 300, 15. doi:10.1017/S1743921313010697

Bi, Y., Jiang, Y., Yang, J., et al. 2015, ApJ, 805, 48. doi:10.1088/0004-637X/805/1/48

Birn, J., Forbes, T. G., & Hesse, M. 2006, ApJ, 645, 732. doi:10.1086/504280

Brueckner, G. E., Howard, R. A., Koomen, M. J., et al. 1995, SoPh, 162, 357. doi:10.1007/BF00733434

Byrne, J. P., Morgan, H., Seaton, D. B., et al. 2014, SoPh, 289, 4545. doi:10.1007/s11207-014-0585-8

Cao, W., Gorceix, N., Coulter, R., et al. 2010, Astronomische Nachrichten, 331, 636. doi:10.1002/asna.201011390

Cao, W., Coulter, R., Gorceix, N., et al. 2010, Proc. SPIE, 7742, 774220. doi:10.1117/12.856616

Chae, J., Denker, C., Spirock, T. J., et al. 2000, SoPh, 195, 333. doi:10.1023/A:1005242832293

Chandra, R., Filippov, B., Joshi, R., et al. 2017, SoPh, 292, 81. doi:10.1007/s11207-017-1104-5

Chen, J. 1996, J. Geophys. Res., 101, 27499. doi:10.1029/96JA02644

Chen, P. F., Innes, D. E., & Solanki, S. K. 2008, A&A, 484, 487. doi:10.1051/0004-6361:200809544

Chen, P. F. & Shibata, K. 2000, ApJ, 545, 524. doi:10.1086/317803

Chen, P. F. 2011, Living Reviews in Solar Physics, 8, 1. doi:10.12942/lrsp-2011-1

Chen, H., Duan, Y., Yang, J., et al. 2018, ApJ, 869, 78. doi:10.3847/1538-4357/aaead1

Cheng, X. & Ding, M. D. 2016, ApJS, 225, 16. doi:10.3847/0067-0049/225/1/16

Cheng, X., Kliem, B., & Ding, M. D. 2018, ApJ, 856, 48. doi:10.3847/1538-4357/aab08d

Chifor, C., Tripathi, D., Mason, H. E., et al. 2007, A&A, 472, 967. doi:10.1051/0004-6361:20077771

Dai, J., Yang, J., Li, L., et al. 2018, ApJ, 869, 118. doi:10.3847/1538-4357/aaedbb

Dai, J., Ji, H., Li, L., et al. 2021, ApJ, 906, 66. doi:10.3847/1538-4357/abcaf4

Devi, P., Joshi, B., Chandra, R., et al. 2020, SoPh, 295, 75. doi:10.1007/s11207-020-01642-y

Engvold, O. 1998, IAU Colloq. 167: New Perspectives on Solar Prominences, 150, 23

Engvold, O. 2015, Solar Prominences, 31. doi:10.1007/978-3-319-10416-4_2

Fang, F. & Fan, Y. 2015, ApJ, 806, 79. doi:10.1088/0004-637X/806/1/79

Filippov, B. 2018, MNRAS, 475, 1646. doi:10.1093/mnras/stx3277

Feynman, J. & Martin, S. F. 1995, J. Geophys. Res., 100, 3355. doi:10.1029/94JA02591

Hernandez-Perez, A., Su, Y., Veronig, A. M., et al. 2019, ApJ, 874, 122. doi:10.3847/1538-4357/ab09ed

Gibson, S. E. & Fan, Y. 2006, ApJL, 637, L65. doi:10.1086/500452

Gibson, S. E. 2018, Living Reviews in Solar Physics, 15, 7. doi:10.1007/s41116-018-0016-2

Gilbert, H. R., Holzer, T. E., Burkepile, J. T., et al. 2000, ApJ, 537, 503. doi:10.1086/309030

Gilbert, H. R., Holzer, T. E., & Burkepile, J. T. 2001, ApJ, 549, 1221. doi:10.1086/319444

Gilbert, H. R., Alexander, D., & Liu, R. 2007, SoPh, 245, 287. doi:10.1007/s11207-007-9045-z

Goode, P. R., Coulter, R., Gorceix, N., et al. 2010, Astronomische Nachrichten, 331, 620. doi:10.1002/asna.201011387

Guo, Y., Xia, C., & Keppens, R. 2016, ApJ, 828, 83. doi:10.3847/0004-637X/828/2/83

Gosain, S., Filippov, B., Ajor Maurya, R., et al. 2016, ApJ, 821, 85. doi:10.3847/0004-637X/821/2/85

Wiegelmann, T., Thalmann, J. K., Inhester, B., et al. 2012, SoPh, 281, 37. doi:10.1007/s11207-012-9966-z

Wiegelmann, T., Inhester, B., & Sakurai, T. 2006, SoPh, 233, 215. doi:10.1007/s11207-006-2092-z

Karpen, J. T. 2015, Solar Prominences, 415, 237. doi:10.1007/978-3-319-10416-4_10

Keppens, R., Nool, M., Toth, G., et al. 2003, Computer Physics Communications, 153, 317. doi:10.1016/S0010-4655(03)00139-5

Keppens, R., Meliani, Z., van Marle, A. J., et al. 2012, Journal of Computational Physics, 231, 718. doi:10.1016/j.jcp.2011.01.020

Porth, O., Xia, C., Hendrix, T., et al. 2014, ApJS, 214, 4. doi:10.1088/0067-0049/214/1/4

Haerendel, G. & Berger, T. 2011, ApJ, 731, 82. doi:10.1088/0004-637X/731/2/82

Hagyard, M. J., Smith, J. B., Teuber, D., et al. 1984, SoPh, 91, 115. doi:10.1007/BF00213618

Hood, A. W. & Priest, E. R. 1981, Geophysical and Astrophysical Fluid Dynamics, 17, 297. doi:10.1080/03091928108243687

Ji, H., Wang, H., Schmahl, E. J., et al. 2003, ApJL, 595, L135. doi:10.1086/378178

Kaifan Ji, Hui Liu, Zhenyu Jin, Zhenhong Shang, Zhenping Qiang, In Journal of Chinese Science Bulletin,2019, https://doi.org/10.1360/N972019-00092.

Kliem, B. & Török, T. 2006, PhRvL, 96, 255002. doi:10.1103/PhysRevLett.96.255002

Kliem, B., Török, T., Titov, V. S., et al. 2014, ApJ, 792, 107. doi:10.1088/0004-637X/792/2/107

Labrosse, N., Heinzel, P., Vial, J.-C., et al. 2010, SSRv, 151, 243. doi:10.1007/s11214-010-9630-6

Lemen, J., Title, A., Akin, D., et al. 2012, SoPh, 275, 17. doi:10.1007/s11207-011-9776-8

Li, L., Zhang, J., Peter, H., et al. 2016, Nature Physics, 12, 847. doi:10.1038/nphys3768

Lin, Y., Engvold, O., Rouppe van der Voort, L., et al. 2005, SoPh, 226, 239. doi:10.1007/s11207-005-6876-3

Lin, Y., Engvold, O., Martin, S., et al. 2008, AGU Spring Meeting Abstracts

Liu, R., Alexander, D., & Gilbert, H. R. 2007, ApJ, 661, 1260. doi:10.1086/513269

Liu, Y., Su, J., Xu, Z., et al. 2009, ApJL, 696, L70. doi:10.1088/0004-637X/696/1/L70

Liu, R., Kliem, B., Török, T., et al. 2012, ApJ, 756, 59. doi:10.1088/0004-637X/756/1/59

Liu, L., Liu, J., Chen, J., et al. 2021, A&A, 648, A106. doi:10.1051/0004-6361/202140277

Mackay, D. H., Karpen, J. T., Ballester, J. L., et al. 2010, SSRv, 151, 333. doi:10.1007/s11214-010-9628-0

Mitra, P. K., Joshi, B., & Prasad, A. 2020, SoPh, 295, 29. doi:10.1007/s11207-020-1596-2

Martin, S. F. 1980, SoPh, 68, 217. doi:10.1007/BF00156861

Martin, S. F. 1998, SoPh, 182, 107. doi:10.1023/A:1005026814076

Monga, A., Sharma, R., Liu, J., et al. 2021, MNRAS, 500, 684. doi:10.1093/mnras/staa2902

Moore, R. L., Sterling, A. C., Hudson, H. S., et al. 2001, ApJ, 552, 833. doi:10.1086/320559

Okamoto, T. J., Tsuneta, S., Berger, T. E., et al. 2007, Science, 318, 1577. doi:10.1126/science.1145447

Pan, H., Liu, R., Gou, T., et al. 2021, ApJ, 909, 32. doi:10.3847/1538-4357/abda4e

Pesnell, W., Thompson, B., & Chamberlin, P. 2012, SoPh, 275, 3

Régnier, S., Walsh, R. W., & Alexander, C. E. 2011, A&A, 533, L1. doi:10.1051/0004-6361/201117381

Parenti, S. 2014, Living Reviews in Solar Physics, 11, 1. doi:10.12942/lrsp-2014-1

Schou, J., Scherrer, P., Bush, R., et al. 2012, SoPh, 275, 229. doi:10.1007/s11207-011-9842-2

Schmieder, B., Török, T., & Aulanier, G. 2008, Exploring the Solar System and the Universe, 1043, 260. doi:10.1063/1.2993658

Schmieder, B., Chandra, R., Berlicki, A., et al. 2010, A&A, 514, A68. doi:10.1051/0004-6361/200913477

Schrijver, C. J. 2007, ApJL, 655, L117. doi:10.1086/511857

Shen, Y., Liu, Y., & Su, J. 2012, ApJ, 750, 12. doi:10.1088/0004-637X/750/1/12

Shen, Y.-D., Liu, Y., & Liu, R. 2011, Research in Astronomy and Astrophysics, 11, 594. doi:10.1088/1674-4527/11/5/009

Sterling, A. C. & Moore, R. L. 2005, ApJ, 630, 1148. doi:10.1086/432044

Sterling, A. C., Moore, R. L., & Freeland, S. L. 2011, ApJL, 731, L3. doi:10.1088/2041-8205/731/1/L3

Tanaka, K. 1991, SoPh, 136, 133. doi:10.1007/BF00151700

Tripathi, D., Gibson, S. E., Qiu, J., et al. 2009, A&A, 498, 295. doi:10.1051/0004-6361/200809801

Tripathi, D., Reeves, K. K., Gibson, S. E., et al. 2013, ApJ, 778, 142. doi:10.1088/0004-637X/778/2/142

Török, T. & Kliem, B. 2005, ApJL, 630, L97. doi:10.1086/462412

Török, T., Kliem, B., & Titov, V. S. 2004, A&A, 413, L27. doi:10.1051/0004-6361:20031691

Török, T., Chandra, R., Pariat, E., et al. 2011, ApJ, 728, 65. doi:10.1088/0004-637X/728/1/65

Turmon, M., Jones, H. P., Malanushenko, O. V., et al. 2010, SoPh, 262, 277. doi:10.1007/s11207-009-9490-y

van Ballegooijen, A. A. & Martens, P. C. H. 1989, ApJ, 343, 971. doi:10.1086/167766

Wang, Y., Su, Y., Shen, J., et al. 2018, ApJ, 859, 148. doi:10.3847/1538-4357/aac0f7

Wang, Y., Ji, H., Warmuth, A., et al. 2020, ApJ, 905, 126. doi:10.3847/1538-4357/abc47a

Xue, Z., Yan, X., Cheng, X., et al. 2016, Nature Communications, 7, 11837. doi:10.1038/ncomms11837

Yan, X. L., Qu, Z. Q., Kong, D. F., et al. 2013, A&A, 557, A108. doi:10.1051/0004-6361/201219032

Yang, J., Hong, J., Li, H., et al. 2020, ApJ, 900, 158. doi:10.3847/1538-4357/aba7c0

Zeng, Z., Qiu, J., Cao, W., et al. 2014, ApJ, 793, 87. doi:10.1088/0004-637X/793/2/87

Zhang, Q. M., Chen, P. F., Xia, C., et al. 2012, A&A, 542, A52. doi:10.1051/0004-6361/201218786

Zhang, Q. M., Ning, Z. J., Guo, Y., et al. 2015, ApJ, 805, 4. doi:10.1088/0004-637X/805/1/4

Zheng, R., Zhang, Q., Chen, Y., et al. 2017, ApJ, 836, 160. doi:10.3847/1538-4357/aa5c38

Zheng, R., Yang, S., Rao, C., et al. 2019, ApJ, 875, 71. doi:10.3847/1538-4357/ab0f3f

Chandra, R., Schmieder, B., Mandrini, C. H., et al. 2011, SoPh, 269, 83. doi:10.1007/s11207-010-9670-9

DeVore, C. R., Antiochos, S. K., & Aulanier, G. 2005, ApJ, 629, 1122. doi:10.1086/431721

Schmieder, B., Mein, N., Deng, Y., et al. 2004, SoPh, 223, 119. doi:10.1007/s11207-004-1107-x